\documentclass[pra,showpacs,superscriptaddress,a4paper,preprint,eqsecnum]{revtex4}
\usepackage{graphicx}
\usepackage{psfrag}

\newcommand{\sn}{\mathrm{sn}}
\newcommand{\cn}{\mathrm{cn}}

\newcommand{\sech}{\mathrm{sech}}
\newcommand{\vb}[1]{{\mbox{\boldmath$#1$}}}
\newcommand{\be}{\begin{equation}}
\newcommand{\ee}{\end{equation}}

\begin{document}

\title{Stationary states of Bose-Einstein condensates in single- and multi-well trapping potentials}
\author{Roberto D'Agosta} 
\affiliation{Dipartimento di Fisica ``E. Amaldi'', Universit\`a di Roma 3,
Via della Vasca Navale 84, Roma 00146 Italy}
\affiliation{Istituto Nazionale per la Fisica della Materia, Unit\`a di Roma 3}
\author{Boris A. Malomed} 
\affiliation{Department of Interdisciplinary Studies, Faculty of Engineering, Tel Aviv University, Tel Aviv 69978, Israel}
\author{Carlo Presilla}
\affiliation{Dipartimento di Fisica, Universit\`a di Roma ``La Sapienza'',
Piazzale A. Moro 2, Roma 00185, Italy}
\affiliation{Istituto Nazionale per la Fisica della Materia, Unit\`a di Roma 1} 
\affiliation{Istituto Nazionale di Fisica Nucleare, Sezione di Roma 1} 
\date{\today}

\begin{abstract}
The stationary solutions of the Gross-Pitaevskii equation can be 
divided in two classes: those which reduce, in the limit of vanishing 
nonlinearity, to the eigenfunctions of the associated Schr\"odinger 
equation and those which do not have linear counterpart.
Analytical and numerical results support an existence condition 
for the solutions of the first class in terms of the ratio between 
their proper frequency and the corresponding linear eigenvalue. 
For one-dimensional confined systems, we show that solutions without 
linear counterpart do exist in presence of a multi-well external 
potential. 
These solutions, which in the limit of strong nonlinearity have the 
form of chains of dark or bright solitons located near the extrema 
of the potential, represent macroscopically excited states of a 
Bose-Einstein condensate and are in principle experimentally 
observable. 
\end{abstract}
\maketitle

\section{Introduction}

Bose-Einstein condensates of gases of alkali atoms confined in 
magnetic or optic traps are effectively described by the 
Gross-Pitaevskii equation (GPE), a Schr\"odinger equation with a local 
cubic nonlinear term which takes into account the interaction among 
the bosons in a mean field approximation \cite{isw}. 
The stationary solutions of the GPE represent macroscopically 
excited states of the condensate and have attracted great 
theoretical interest \cite{yyb,kat}.
The existence of some of these states have been also demonstrated 
in recent experiments.
Vortices have been observed in two- \cite{matthews}
or one-component \cite{madison} condensates.
Phase engineering optical techniques have allowed to generate dark 
solitons in atomic gases with positive scattering length 
\cite{burger,denschlag}.

The excited states observed in \cite{matthews,madison,burger,denschlag}
have linear counterpart, i.e. they are stationary solutions
of the GPE which reduce, in the limit of vanishing 
nonlinearity, to the eigenfunctions of the associated Schr\"odinger 
equation \cite{dmp}.
However, the GPE can admit also a class of stationary states
without linear counterpart. 
These solutions appear for a sufficiently large value of the 
nonlinearity whenever the system is confined in an external 
multi-well potential \cite{dp}.

In this paper we review the general properties of the stationary 
solutions of the GPE.
In Section \ref{with} we discuss an existence condition for the
solutions with linear counterpart in terms of the ratio between 
their proper frequency and the corresponding linear eigenvalue. 
We also show that in the limit of strong nonlinearity these solutions
assume the shape of chains of dark or bright solitons depending 
on the repulsive or attractive nature of the interaction.
In Section \ref{without} we generalize the asymptotic shape of the states
with linear counterpart to find a new kind of solutions of the GPE.
These correspond to solitons located near the extrema of the external
potential in a way which generally breaks the symmetry of the system. 
We consider the particular case of a symmetric double-well
in Section \ref{twowell} and describe all the zero-, one-, 
and two-soliton solutions of this system.
In Section \ref{birth} we show an example of 
how the solutions without linear counterpart are generated 
by increasing the nonlinearity. 
The stability properties of the stationary solutions in view of
a possible experimental observation are pointed out in Section 
\ref{conclusions}. 

\section{Solutions with linear counterpart}\label{with}

The stationary solutions of the GPE are defined as
\be
\Psi_\mu(\vb{x},t)=e^{-\frac{i}{\hbar}\mu t} \psi_\mu(\vb{x}),
\ee
where $\mu$ is called chemical potential, 
and determined by the equation
\be
-\frac{\hbar^2}{2m}\nabla^2\psi_\mu(\vb{x})+
U_0|\psi_\mu(\vb{x})|^2\psi_\mu(\vb{x})
+V(\vb{x})\psi_\mu(\vb{x})=\mu\psi_\mu(\vb{x})
\ee
with the normalization condition $\Vert \psi_\mu \Vert^2 = N[\psi_\mu]$.
For later use we note that the stationary solutions are also 
critical points of the grand-potential functional
\begin{eqnarray}
\Omega[\psi] &=&
\int\left[ \frac{\hbar^2}{2m}|\nabla\psi(\vb{x})|^2
+\frac{U_0}{2}|\psi(\vb{x})|^4
+\left( V(\vb{x})- \mu \right) |\psi(\vb{x})|^2
\right] d\vb{x} \nonumber \\
&=& E[\psi]-\mu N[\psi]. 
\end{eqnarray}

Let us consider stationary solutions $\psi_{\mu n}$ which reduce for 
$U_0 \to 0$ to the eigenfunctions $\phi_n$ of the Schr\"odinger equation
\be
-\frac{\hbar^2}{2m}\nabla^2\phi_n(\vb{x})+V(\vb{x})\phi_n(\vb{x})
={\cal E}_n\phi_n(\vb{x}).
\ee
This is the linear limit of the GPE which can be obtained, 
for $U_0$ fixed,
by varying the norm of the solutions, i.e. the chemical potential $\mu$.
In fact, by substituting in the GPE, 
$\psi_{\mu n}(\vb{x})=\sqrt{N_n(\mu)}\chi_{\mu n}(\vb{x})$, 
with $N_n(\mu)=\Vert \psi_{\mu n} \Vert^2$ and 
$\Vert \chi_{\mu n}\Vert^2=1$, 
we get
\be
-\frac{\hbar^2}{2m}\nabla^2\chi_{\mu n}(\vb{x})+
U_0 N_n(\mu) |\chi_{\mu n}(\vb{x})|^2\chi_{\mu n}(\vb{x})+
V(\vb{x})\chi_{\mu n}(\vb{x})=\mu \chi_{\mu n}(\vb{x}).
\ee
For $N_n(\mu)$ small, 
the nonlinear term can be neglected and $\chi_{\mu n} \simeq \phi_n$. 
Moreover, we have
\be
\mu \simeq {\cal E}_n+U_0 N_n(\mu) ||\phi_n^2||^2,  
\ee
i.e. for $N_n(\mu) \to 0$ the chemical potential tends to ${\cal E}_n$ 
form above or below depending on the sign of $U_0$.

The above considerations suggest the following existence conjecture.
{\sl
For $U_0>0$ $(U_0<0)$, solutions with linear limit 
$\psi_{\mu n} \simeq \sqrt{N_n(\mu)}\phi_n$ 
exist only if $\mu>{\cal E}_n$ $(\mu<{\cal E}_n)$.
Moreover $N_n(\mu)\to 0$ for $\mu \to {\cal E}_n$}.
This conjecture can be i) proved by a theorem in the case $n=0$,
ii) explicitly verified in the solvable case of a one-dimensional 
system confined in a box, and iii) supported by numerical results 
for multidimensional systems with different potentials \cite{dmp}.

Here, we illustrate point ii) whose results are useful also for
the subsequent discussion on general external potentials.
Let us consider the case of a one-dimensional system confined in a 
box extending from $-L/2$ to $L/2$. 
For $U_0>0$, the Jacobi elliptic functions 
\begin{equation}
\psi_{\mu n}(x)=A~\sn \left(\left. 2(n+1)\mbox{K}(p)\left(\frac{x}{L}+\frac{1}{2}\right)\right|p\right),
\label{sol1d}
\end{equation}
where 
\begin{equation}
\mbox{K}(p)=\int_0^{\frac{\pi}{2}}\frac{1}{\sqrt{1-p\sin^2\theta}}d\theta
\end{equation}
is the complete elliptic integral of the first kind with modulus 
$p\in [0,1]$, and  $n=0,1,2,\ldots$, 
solve the GPE under the conditions 
\begin{eqnarray}
A^2&=&\frac{\hbar^2}{mU_0L^2}~p~(2(n+1)\mbox{K}(p))^2,\\
\mu &=& \frac{\hbar^2}{mL^2}~\frac{p+1}{2}~(2(n+1)\mbox{K}(p))^2.
\label{mu1}
\end{eqnarray}
Since $\mbox{K}(p)$ increases monotonously from $\mbox{K}(0)=\pi/2$, 
for a given $n$ Eq. (\ref{mu1}) has solution only if 
\begin{equation}
\mu\geq \mathcal{E}_n\equiv \frac{(n+1)^2\pi^2\hbar^2}{2mL^2}.
\end{equation}
This complies with the conjecture formulated above.

For $U_0<0$, the solutions of the GPE in the box are of the form
\begin{equation}
\psi_{\mu n}(x)=
A~\cn\left(\left. 2(n+1)\mbox{K}(p)\left(\frac{x}{L}+
\frac{1}{2}\right)+\mbox{K}(p)\right|p\right)
\label{sol1d_neg}
\end{equation}
with the conditions
\begin{eqnarray}
A^2&=&-\frac{\hbar^2}{mU_0L^2}~p~(2(n+1)\mbox{K}(p))^2,\\
\mu&=& \frac{\hbar^2}{mL^2}~\frac{1-2p}{2}~(2(n+1)\mbox{K}(p))^2.
\label{mu2}
\end{eqnarray}
Since $(1-2p)\mbox{K}(p)$ decreases monotonously for $p\in [0,1]$, 
the $n$-node solution exists only if $\mu\leq \mathcal{E}_n$ in agreement
with the conjecture.

For both $U_0 >0$ and $U_0<0$,
the stationary solutions of the GPE in the box reduce in the linear limit 
to the well known Schr\"odinger eigenfunctions
\begin{equation}
\frac{1}{\sqrt{N_n(\mu)}}\psi_{\mu n}(x)\stackrel{\mu \to \mathcal{E}_n}{\longrightarrow}\sqrt{\frac{2}{L}}\sin\left[\left(\frac{x}{L}+\frac{1}{2}\right)(n+1)\pi\right].
\label{linlim}
\end{equation}  
In the opposite limit of strong nonlinearity, we obtain chains of 
solitons.
For $U_0>0$ and $\mu \gg \mathcal{E}_n$, 
we get dark soliton solutions 
\begin{eqnarray}
\psi_{\mu n}(x)&\stackrel{\mu \gg \mathcal{E}_n}{\longrightarrow}&\sqrt{\frac{\mu}{U_0}}~\prod_{k=0}^{n+1}\tanh \left(\frac{\sqrt{m\mu}}{\hbar}(x-x_k)\right),
\end{eqnarray}   
with solitons centered at 
$x_k=\left(-\frac{1}{2}+\frac{1}{n+1}k\right)L$.
For $U_0<0$ and $-\mu \gg \mathcal{E}_n$, 
we have bright soliton solutions 
\begin{eqnarray}
\psi_{\mu n}(x)&\stackrel{-\mu \gg \mathcal{E}_n}{\longrightarrow}&\sqrt{\frac{2 \mu}{U_0}}~\sum_{k=0}^{n}(-1)^k \sech \left(\frac{\sqrt{-2 m\mu}}{\hbar}(x-x_k)\right),
\end{eqnarray}
with solitons located in
$x_k=\left[-\frac{1}{2}+\frac{1}{n+1}\left(k+\frac{1}{2}\right)\right]L$.

\section{Solutions without linear counterpart}\label{without}

The soliton chains obtained in Sec. \ref{with} as stationary solutions 
of the GPE in a box can be generalized in the case of a potential $V$ 
of arbitrary shape.
Let us consider first the case $U_0>0$. 
For $\mu \to \infty$,
the repulsive interaction tends to delocalize the solutions so that 
a Thomas-Fermi approximation holds, i.e. we can neglect the
gradient term in the GPE 
\be
U_0|\psi(\vb{x})|^2\psi(\vb{x})+V(\vb{x})\psi(\vb{x})
\simeq \mu\psi(\vb{x}).
\ee
Therefore, the GPE has always the solution
$$
\psi_{\mu 0}(\vb{x})=\left\{
\begin{array}{ll}
\sqrt{\left(\mu-V(\vb{x})\right)/U_0}& \mu>V(\vb{x})\\
0 & \mu<V(\vb{x})
\end{array}  
\right. 
$$
and, in the one-dimensional case, also $n$-node solutions of the form
\be
\psi_{\mu n}(x)=\psi_{\mu 0}(x)\prod_{k=1}^{n}\tanh 
\left(\frac{\sqrt{m\mu}}{\hbar}(x-x_k)\right),
\label{dsc}
\ee
provided that the solitons do not overlap, i.e. 
$|x_{k+1}-x_k| \gg \hbar/\sqrt{m\mu}$. 

In the attractive case $U_0<0$, for $\mu \to -\infty$
the mean field density tends to be localized and the linear potential
in the GPE can be neglected with respect to the cubic one
\be
-\frac{\hbar^2}{2m}\nabla^2\psi(\vb{x})+
U_0|\psi(\vb{x})|^2\psi(\vb{x})
\simeq \mu\psi(\vb{x}). 
\ee
In the one-dimensional case, this equation has solutions of the form
\be
\psi_{\mu n}(x)=\sqrt{\frac{2 \mu}{U_0}}~\sum_{k=0}^{n} s_k~ \sech 
\left(\frac{\sqrt{-2 m\mu}}{\hbar}(x-x_k)\right),
\label{bsc}
\ee
provided that $|x_{k+1}-x_k| \gg \hbar/\sqrt{-2m\mu}$.
Note that for the corresponding solutions with linear counterpart 
we must have $s_k=(-1)^k$.
However, all sign combinations $s_k=\pm 1$ are possible in general. 

In order the dark- and bright-soliton chains (\ref{dsc}) and (\ref{bsc})
to be asymptotic 
(for $\mu \to \infty$ or $\mu \to - \infty$, respectively)
solutions of the GPE, the soliton centers must be extremal points
of the corresponding grand-potential $\Omega(\{x_k\})$.
Depending on the shape of the external potential, stationary
solutions without linear counterpart may arise.
Consider, for instance, the one-soliton solutions.
In the strongly nonlinear limit, the soliton width, 
$\hbar/\sqrt{m\mu}$ for $U_0>0$
($\hbar/\sqrt{-2m\mu}$ for $U_0<0$), vanishes and 
$\Omega(x_1) \sim {\rm const} - V(x_1)$
($\Omega(x_0) \sim {\rm const} + V(x_0)$ for $U_0<0$).
If the external potential $V(x)$ is single-well,  
$\Omega(x_1)$ or $\Omega(x_0)$ will have only one extremum. 
The corresponding solution necessarily reduces
in the linear limit to the 1-node, for $U_0>0$, or 0-node, for $U_0<0$,  
Schr\"odinger eigenfunction.
However, if the external potential is multi-well, several one-soliton
solutions are possible.
Some of them will not have linear counterpart, i.e. they disappear 
when the linear limit is approached. 

The approximate dark- or bright-soliton solutions now introduced 
make possible a numerical search for the GPE stationary solutions 
for any value of the nonlinearity.
In fact, stationary solutions of a PDE can be found numerically 
by using a relaxation algorithm which converges to the solution 
``closest'' to a given input function.
Thus, only solutions of which a good approximation is known can be found.
In practice, the numerical search of stationary solutions can be 
organized in the following way: 
\begin{itemize}
\item[-] choose a sufficiently large $|\mu|$ and consider an asymptotic 
solution of the form of a dark- or bright-soliton chain;
\item[-] determine the soliton centers $x_k$ by extremizing the 
corresponding grand-potential $\Omega(\{x_k\})$;
\item[-] use this approximate solution as input function in the 
relaxation algorithm;
\item[-] use the obtained output solution as input function in a new
relaxation with a smaller value of $|\mu|$;
\item[-] repeat the last step until the linear region is reached.
\end{itemize}

\section{Symmetric double-well system}\label{twowell}

As an example of the general approach outlined in the previous Section, 
now we determine all the zero-, one- and two-solitons
solutions for a system confined in a symmetric double-well potential.
We choose 
\be
V(x)=m^2 \gamma^4 x^4-m\omega^2 x^2+\frac{\omega^4}{4 \gamma^4}
\ee
which has two minima in $x=\pm x_m$, where 
$x_m \equiv \sqrt{\omega^2/2 m \gamma^4}$, and a maximum at $x=0$.

Zero-soliton solutions exist only for the repulsive GPE.
For $\mu$ sufficiently large, just one node-less state is possible.
This state extends over the entire double well and is given by 
$\psi_{\mu 0}(x)=\sqrt{\left(\mu-V(x)\right)/U_0}$
for $x$ such that $V(x)<\mu$ and 
$\psi_{\mu 0}(x)=0$ otherwise.
If $\mu$ is smaller than the barrier height 
$\omega^4/4 \gamma^4$, the above solution still exists 
and eventually approaches the ground state of the Schr\"odinger
equation for the double-well.
However, two other possibilities appear.
We can have a state localized in the left well that joins with the 
identically vanishing solution in the right well and viceversa.
These states break the symmetry of $V(x)$ and do not have linear 
counterpart.

One-soliton solutions are described by Eq. (\ref{dsc}) with $n=1$ in the
repulsive case and Eq. (\ref{bsc}) with $n=0$ in the attractive one.
The corresponding grand-potential becomes a function of the soliton 
centers $x_1$ or $x_0$, respectively.
As shown in Fig. 1, for $|\mu|$ sufficiently large we have  
$\Omega(x_1) \sim {\rm const} -V(x_1)$ and
 $\Omega(x_0) \sim {\rm const} +V(x_0)$.
\begin{figure}
\psfrag{x1/xm}[][][0.9]{$x_1/x_m$}
\psfrag{x0/xm}[][][0.9]{$x_0/x_m$}
\psfrag{O(x1)/|O(0)|}[][][0.9]{$\Omega(x_1)/|\Omega(0)|$}
\psfrag{O(x0)/O(0)}[][][0.9]{$\Omega(x_0)/\Omega(0)$}
\includegraphics[width=11cm]{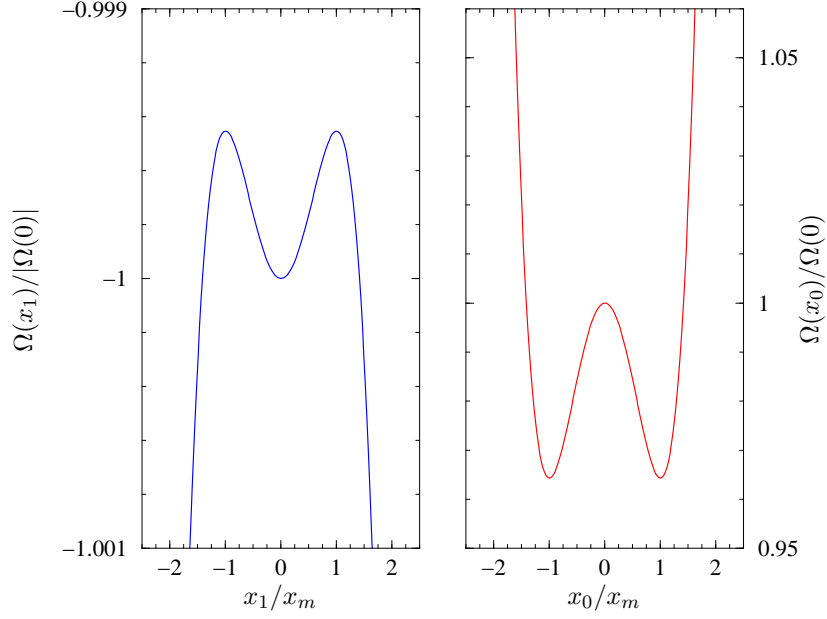}
\caption{ Grand potential $\Omega$ as a function of the dark soliton 
center $x_1$ (left panel) and the bright soliton center $x_0$ 
(right panel) for the one-soliton solutions.}
\end{figure}
In both cases we have three one-soliton solutions corresponding to the 
three extrema of the external potential.
The soliton may be found in the maximum
or in one of the two minima of the double-well.
For $U_0>0$, the solution of the first kind reduces in the linear limit 
to the anti-symmetric Schr\"odinger eigenstate with a single node.
For $U_0<0$, the solution with a bright soliton in the 
maximum of the double-well, even respecting the symmetry
of the potential, disappears in the linear limit.    
The two solutions with the soliton, dark or bright, in one of the 
minima $\pm x_m$ of the double-well
break the symmetry of $V(x)$ and do not have linear counterpart.

In the repulsive case, two-soliton solutions are described by Eq. 
(\ref{dsc}) with $n=2$ and the grand-potential becomes the 
two-variable function $\Omega(x_1,x_2)$ whose contour plot is 
shown in Fig. 2. 
\begin{figure}   
\psfrag{x1/xm}[][][0.9][90]{$x_1/x_m$}
\psfrag{x2/xm}[][][0.9]{$x_2/x_m$}
\includegraphics[width=13.0cm]{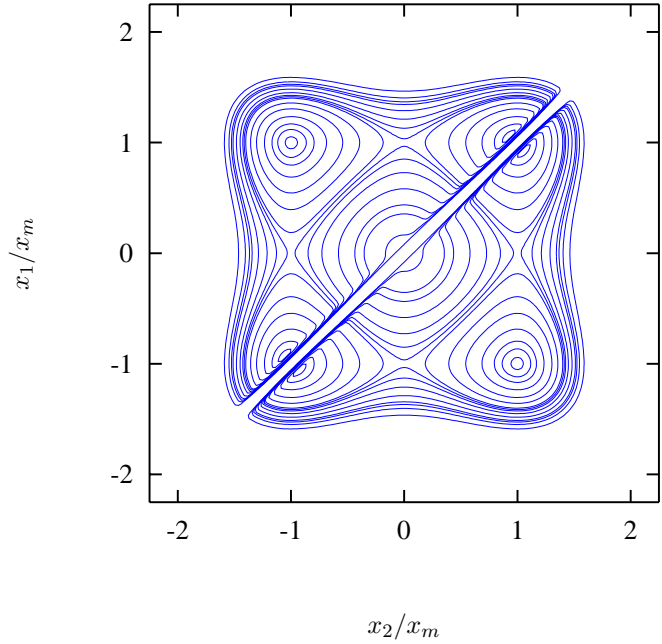}
\caption{
Contour plot of $\Omega(x_1,x_2)$ for the two-soliton solution
in the repulsive case.}
\end{figure}
When the distance between the soliton centers is much larger 
than their width, we have
$\Omega(x_1,x_2) \simeq \Omega(x_1)+\Omega(x_2)$.
In the region $x_1 < x_2$, $\Omega$ has a maximum in 
$(-x_m,x_m)$ and two saddle points in $(0,x_m)$ and $(-x_m,0)$.
We assume that $x_m \gg \hbar/\sqrt{m \mu}$.
The stationary solution corresponding to the maximum of $\Omega$ 
has linear counterpart, namely the symmetric Schr\"odinger eigenstate 
with two nodes in the double-well minima.
Those corresponding to the two saddle points
break the symmetry of $V(x)$ and must disappear in the linear limit.

Other extrema of $\Omega$ can be found
when the centers of the two dark solitons are into the same well.
As shown in Fig. 2, $\Omega$ has two maxima in
$(x_m-\delta,x_m+\delta)$ and $(-x_m-\delta,-x_m+\delta)$
with $2\delta \gtrsim \hbar/\sqrt{m\mu}$. 
The corresponding solutions 
break the symmetry of $V(x)$ and do not have linear counterpart.

In the attractive case, the bright solitons solutions are given 
by Eq. (\ref{bsc}). 
Since the GPE is invariant under a global phase change,
if we restrict to real solutions for $n=1$ (two solitons) 
we have the following two possibilities
\begin{eqnarray}
\psi_{\mu 1}^\pm(x) = \sqrt{\frac{2 \mu}{U_0}}~
\left[
\mbox{sech} \left(\frac{\sqrt{-2 m\mu}}{\hbar}(x-x_0)\right)
\pm 
\mbox{sech} \left(\frac{\sqrt{-2 m\mu}}{\hbar}(x-x_1)\right)
\right].
\label{2bs}
\end{eqnarray}
The contour plots of the functions $\Omega^\pm(x_0,x_1)$ 
obtained by inserting these 
expressions in $\Omega[\psi]$ are shown in Fig. 3.
\begin{figure}   
\psfrag{x1/xm}[][][0.9]{$x_1/x_m$}
\psfrag{x0/xm}[][][0.9][90]{$x_0/x_m$}
\includegraphics[width=13.0cm]{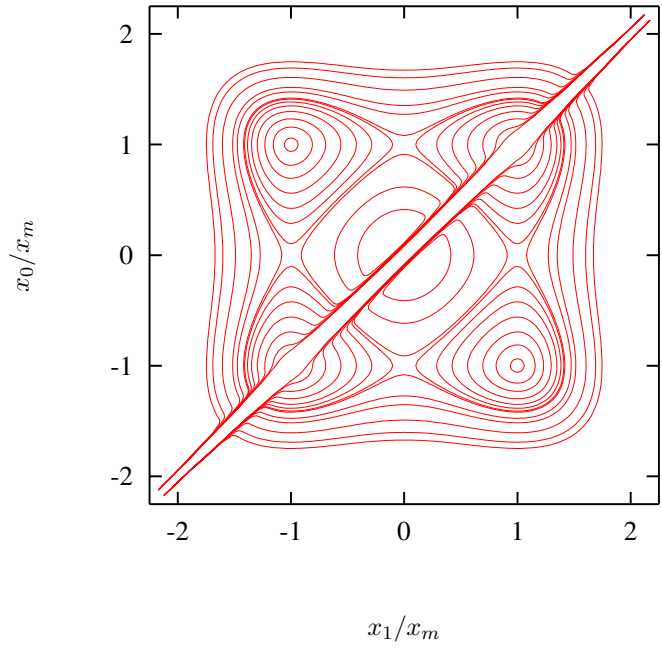}
\includegraphics[width=13.0cm]{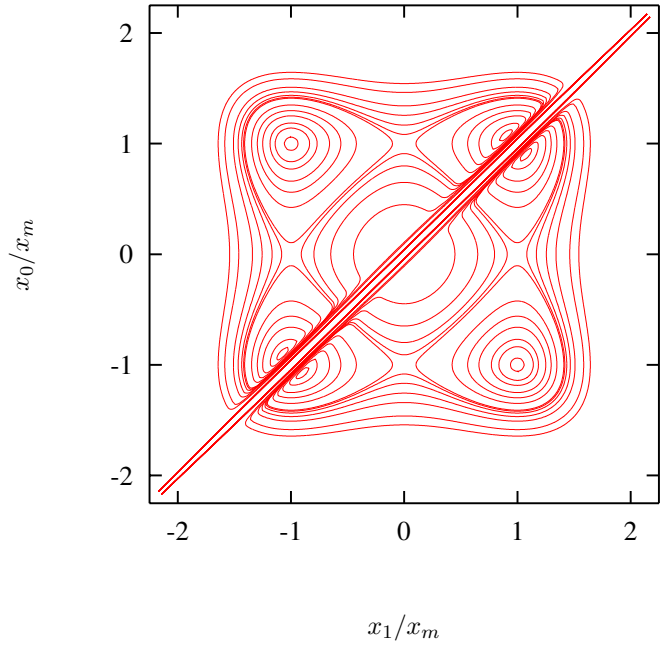}
\caption{
Contour plot of $\Omega^+(x_0,x_1)$ (left) and 
$\Omega^-(x_0,x_1)$ (right) for the two-soliton solution
in the attractive case. The grand potential $\Omega^\pm$ is
evaluated for the two possible real states (\protect\ref{2bs}).}
\end{figure}
In analogy with the repulsive case, for both $\Omega^+$ and $\Omega^-$ 
we have a minimum in $(-x_m,x_m)$ and two saddle points in 
$(0,x_m)$ and $(-x_m,0)$.
The stationary states corresponding to the minimum of 
$\Omega^\pm(x_0,x_1)$ have as linear counterpart the 
lowest-energy symmetric and anti-symmetric Schr\"odinger eigenstates
of the double well. 
Those corresponding to the two saddle points 
break the symmetry of $V(x)$ and do not have linear counterpart.

The functions $\Omega^+$ and $\Omega^-$ have different behavior
when both the soliton centers are inside
the same well, i.e. for $x_0 \sim x_1 \sim \pm 1$.
In fact, $\Omega^+$ does not present new extrema 
while $\Omega^-$, in analogy with the repulsive case, has two minima in 
$(x_m-\delta,x_m+\delta)$ and $(-x_m-\delta,-x_m+\delta)$
with $2\delta \gtrsim \hbar/\sqrt{-2 m \mu}$. 
The corresponding solutions 
break the symmetry of $V(x)$ and do not have linear counterpart.

An example of stationary states without linear counterpart
is shown in Fig. 4. 
\begin{figure}   
\psfrag{x/xm}[][][0.9]{$x/x_m$}
\psfrag{|psi|}[][][0.9]{$\left|\psi_{\mu n}(x)\right|^2$}
\includegraphics[width=9.0cm]{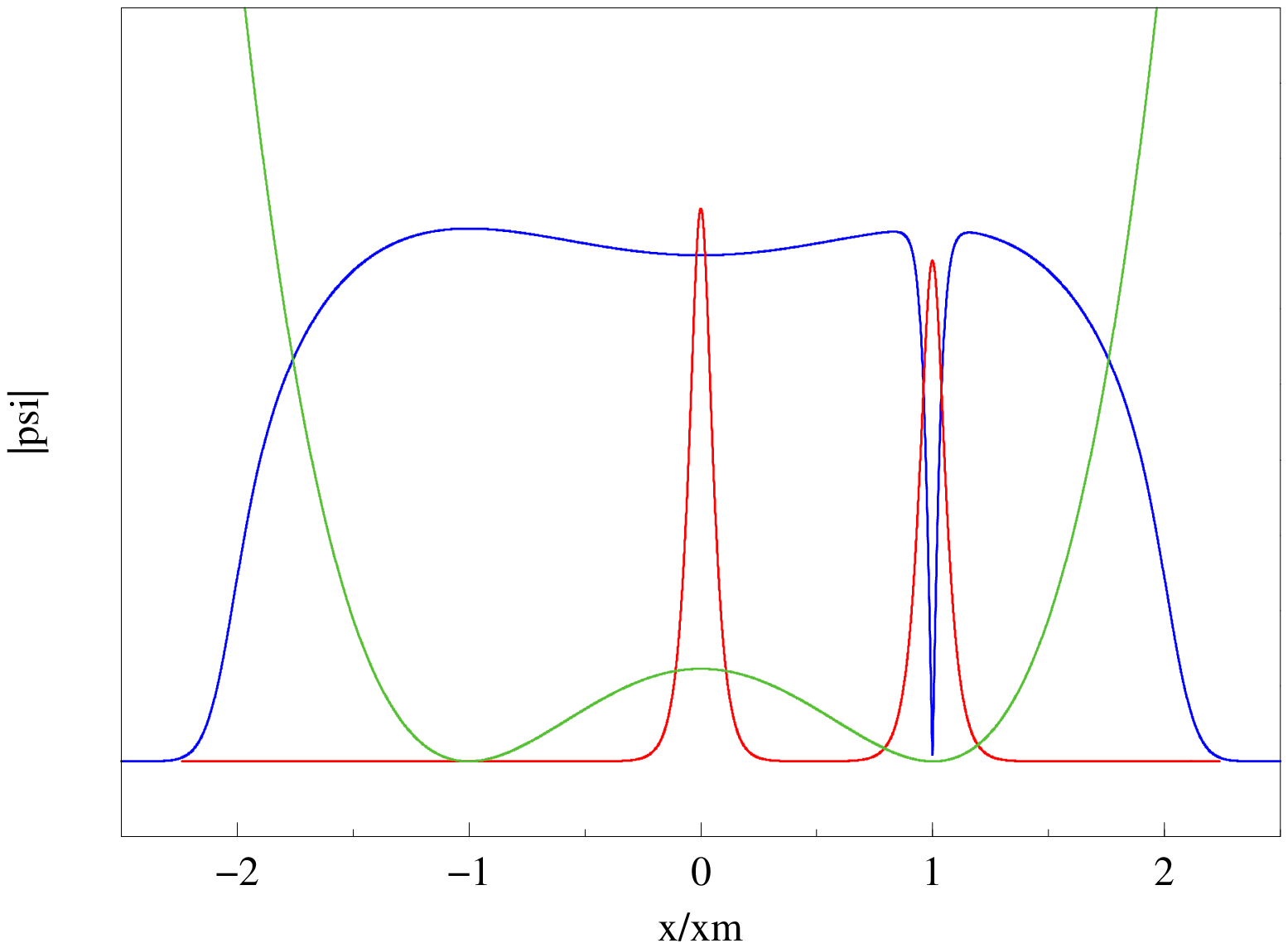}
\caption{
Density of the stationary states 
$\psi_{\mu 1}(x)$ repulsive and $\psi^+_{\mu 1}(x)$ attractive.
For comparison, we show also the double-well potential $V(x)$.}
\end{figure}
These states have been calculated numerically with the procedure 
outlined in Section \ref{without}.  
The parameters used are those of a realistic condensate:
$m=3.818\times 10^{-26}~ {\rm Kg}$,
$\omega=12.75~ {\rm Hz}$,
$\gamma=10^9~ {\rm Kg^{-\frac{1}{4}}m^{-\frac{1}{2}}s^{-\frac{1}{2}}}$,
$U_0= 1.1087\times 10^{-41}~ {\rm Jm}$.
With these values, the distance between the double-well minima 
is $2x_m \simeq 92$ $\mu$m.

\section{Birth of stationary solutions}\label{birth}

To understand how the GPE solutions without linear counterpart arise 
by departing from the linear limit, consider the following example.
We assume a low tunneling regime between the two wells
of the potential of Section \ref{twowell}, 
i.e. $\omega^3/\hbar\gamma^4\gg 1$, and set
\be
\psi(x)=\sqrt{N}\left[a_0\phi_0(x+x_m)+ b_0\phi_0(x-x_m)\right],
\qquad a_0^2+ b_0^2=1,
\label{e1}
\ee
where $\phi_n(x)$ is the $n$-th eigenfunction of the Schr\"odinger 
problem with harmonic potential $\frac12 m (2\omega)^2 x^2$.
Since $\psi$ is already normalized to $N$,
for it to be a stationary solution of the GPE we have to extremize 
the energy functional $E[\psi]=\Omega[\psi]+\mu N[\psi]$.
Up to exponentially small terms, we have
\begin{eqnarray}
E(b_0) &=& N\hbar\omega \left[
\left(1+\frac{3}{16}\frac{\hbar\gamma^4}{\omega^3}\right)+
e^{-\frac{\omega^3}{\hbar\gamma^4}}
\left(1+\frac{3}{8}\frac{\hbar\gamma^4}{\omega^3}-
\frac{3}{2}\frac{\omega^3}{\hbar\gamma^4}\right) b_0\sqrt{1-b_0^2} \right.
\nonumber \\&& \left.
+ \frac{N U_0}{2\sqrt{\pi}}\sqrt{\frac{m}{\hbar^3\omega}}
\left( 1+2 b_0^4-2b_0^2 \right) \right].
\end{eqnarray}
This can be rewritten, up to a constant, as
\be
E(b_0) \sim b_0\sqrt{1-b_0^2} +
\mbox{sign}(U_0) \frac{N}{N_0} \left( 1+2 b_0^4-2b_0^2 \right),
\ee
where 
\be
N_0=  3\sqrt{\pi} \frac{\omega^3}{\hbar \gamma^4}
e^{-\frac{\omega^3}{\hbar \gamma^4}} 
\sqrt{\frac{\hbar^3\omega}{mU_0^2}}.
\ee
The behavior of $E(b_0)$ for different values of the ratio $N/N_0$
is shown in Fig. 5.
\begin{figure}   
\psfrag{N1}[c][c][0.9]{$N/N_0=0$}
\psfrag{N2}[c][c][0.9]{$N/N_0=0.8$}
\psfrag{N3}[c][c][0.9]{$N/N_0=2$}
\psfrag{N4}[c][c][0.9]{$N/N_0=20$}
\psfrag{E(b0)}[c][c][0.9]{$E(b_0)$}
\psfrag{b0}[c][c][0.9]{$b_0$}
\psfrag{1}[c][c][0.9]{$1$}
\psfrag{0}[c][c][0.9]{$0$}
\psfrag{-1}[c][c][0.9]{$-1$}
\includegraphics[width=12cm]{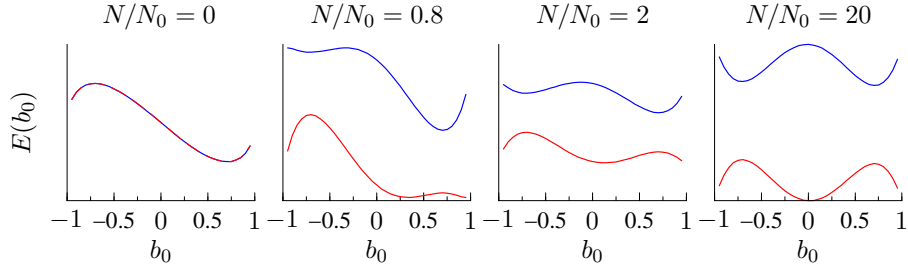}
\caption{Energy of the state (\protect\ref{e1}) as a function of
the parameter $b_0$ for different values of $N/N_0$. 
The upper curve corresponds to the case $U_0<0$,
the lower one to $U_0>0$. For $N=0$ the two curves coincides.}
\end{figure}
For $N \ll N_0$, $E(b_0)$ has a minimum for $b_0=2^{-\frac12}$ 
and a maximum for $b_0=-2^{-\frac12}$. 
These extrema correspond to the lowest energy 
symmetric and anti-symmetric linear states.
If $U_0>0$, for $N \simeq N_0$ the maximum at 
$b_0=-2^{-\frac12}$ bifurcates in a minimum and a maximum. 
Further increasing $N$, the latter moves toward $b_0=0$.
This describes the birth of the state without linear counterpart 
with 0 dark solitons and its localization into the left well. 
If $U_0<0$, the behavior of $E(b_0)$ is similar with maxima and 
minima exchanged. 
In this case, the bifurcation of the minimum at $b_0=2^{-\frac12}$
and its move to $b_0=0$
describes the birth of the state with 1 bright soliton which localizes
into the left well.

\section{Conclusions}\label{conclusions}

We have shown that in presence of an external potential a 1-D GPE
can admit stationary solutions without linear counterpart.
Their existence is strictly connected to the multi-well nature 
of the potential.
In the double well example illustrated here, these solutions disappear
in the limit $\omega \to 0$ when the potential assumes the shape 
of a single quartic well.
For a piece-wise constant double-well, the stationary states here 
discussed analytically only in the limit of strong nonlinearity can be 
obtained in terms of Jacobi elliptic functions for any number of 
particles in the condensate. 

In \cite{dp} we have also investigated the stability of the stationary 
states of the GPE under different points of view. 
The results indicate that the soliton-like states, with and without 
linear counterpart, are sufficiently stable on the typical time scales
of a BEC experiment.
By introducing proper perturbations of the stationary states,
a soliton dynamics could also be observed.

\end{document}